\documentclass[12pt]{article}
\usepackage{epsfig,psfig,doublespace,times,alltt,xspace,subfigure,multicol}

\usepackage{float}
\usepackage[rflt]{floatflt}

\renewcommand{\baselinestretch}{1.5}
\def\floatpagefraction{0.5}
\begin{document}

\newcommand{\topN}{Top-\emph{N}~}
\newcommand{\K}{\emph{k}~}
\newcommand{\ud}{\mathrm{d}}

\newcommand{\vsp}[1]{\renewcommand{\baselinestretch}{#1}}
\newcommand{\defspacing}{\vsp{1.5}}

\newcommand{\eg}{e.g.,\xspace}
\newcommand{\ie}{i.e.,\xspace}

\newcommand{\oneurl}[1]{\texttt{\small #1}}
\newtheorem{kernel-theorem}{Theorem}[section]

\pagestyle{plain}



\title{Supporting Exploratory Queries in \\
Database-Centric Web Applications}

\author{
\begin{tabular}{c c c c}
Abhijit~Kadlag$^\dagger$&
Amol~Wanjari$^\dagger$&
Juliana~Freire$^\ddagger$&
Jayant~R.~Haritsa$^\dagger$
\end{tabular}
\\
\small
\begin{tabular}{c c c}
$^\dagger$Computer Science \& Automation & \hspace*{0.2in} & $^\ddagger$Computer Science \& Engineering \\
Indian Institute of Science &  & OGI/OHSU \\
Bangalore~560012, India & & Beaverton, Oregon 97006, USA 
\end{tabular}
}

\date{}

\maketitle

\setcounter{page}{1}

\begin{abstract}
Users of database-centric Web applications, especially in the e-com\-merce
domain, often resort to exploratory ``trial-and-error'' queries since
the underlying data space is huge and unfamiliar, and there are several
alternatives for search attributes in this space.  For example, scouting
for cheap airfares typically involves posing multiple queries, varying
flight times, dates, and airport locations.  Exploratory queries are
problematic from the perspective of both the user and the server. For
the database server, it results in a drastic reduction in effective
throughput since much of the processing is duplicated in each successive
query. For the client, it results in a marked increase in
response times, especially when accessing the service through wireless
channels.

In this paper, we investigate the design of automated techniques to
minimize the need for repetitive exploratory queries. Specifically, we
present SAUNA, a server-side query relaxation algorithm that, given the
user's initial range query and a desired cardinality for the answer set,
produces a relaxed query that is expected to contain the required number
of answers.  The algorithm incorporates a range-query-specific distance
metric that is weighted to produce relaxed queries of a desired shape
(\eg aspect ratio preserving), and utilizes multi-dimensional 
histograms for query size estimation.  A detailed performance evaluation
of SAUNA over a variety of multi-dimensional data sets indicates that
its relaxed queries can significantly reduce the costs associated with
exploratory query processing.
\end{abstract}

\section{Introduction}
\label{sec:intro}

An increasing number of Web applications are utilizing \emph{database
engines} as their backend information storage system. In fact, a
recent survey~\cite{deepweb} states that more than 200,000 Web sites
generate content from databases containing 7500 terabytes of
information, and they receive 50\% more monthly traffic than other
sites.

Users of database-centric Web applications, especially in the
e-commerce domain, often resort to exploratory ``trial-and-error''
queries since the underlying data space is huge and unfamiliar,
and there are several alternatives for search attributes in this
space~\cite{eureka}.  Consider, for example, the query interface
provided at Travelocity~\cite{www:travelocity}, a popular Web site for
travel planning.  Here, for each itinerary, users must select origin and
destination airports, departure and return times, departure and return
dates, and may optionally select airlines. Faced with this environment,
users often pose a \emph{sequence} of \emph{range queries} while scouting
for cheap airfares.
For example, the first query could be:

{\small
\begin{alltt}
\hspace*{0.3in} SELECT * FROM FLIGHTS 
\hspace*{0.3in} WHERE DepartureTime BETWEEN \emph{10.00 A.M.} AND \emph{11.00 A.M.} AND 
\hspace*{0.3in} DepartureDate BETWEEN \emph{09-11-2003 AND 09-12-2003} AND
\hspace*{0.3in} Origin = "LAX" AND Destination = "JFK" AND Class = "ECONOMY".
\end{alltt}
}

\noindent
and if the result for this query proves to be unsatisfactory,
it is likely to be followed by 

{\small
\begin{alltt}
\hspace*{0.3in}SELECT * FROM FLIGHTS 
\hspace*{0.3in}WHERE DepartureTime BETWEEN \emph{\textbf{08.00 A.M.}} AND \emph{\textbf{12.00 A.M.}} AND 
\hspace*{0.3in}DepartureDate BETWEEN \emph{09-11-2003} AND \emph{\textbf{09-13-2003}} AND
\hspace*{0.3in}Origin = "LAX" AND Destination = "JFK" AND Class = "ECONOMY".
\end{alltt}
}


\noindent
and so on, until a satisfactory result set is obtained.

Such trial-and-error queries are undesirable from the
perspective of both the user and the database server.  For the server,
it results in a drastic reduction in effective throughput since much of
the processing is duplicated in each successive query. For the
client, it results in a marked increase in response times, as well as
frustration from having to submit the query repeatedly.  The problem is
compounded for users who access the Web service through a handheld device
(PDA, smart-phone, etc.) due to the high access latencies, cumbersome
input mechanisms, and limited power supply.

\subsection*{Too Few Answers}
A primary reason for the user dissatisfaction that results in repetitive
queries is the \emph{cardinality} of the answer set -- the Web service
may return \emph{no} or insufficiently few answers, and worse, give
no indication of how to alter the query to provide the desired number
of answers~\cite{eureka}.   (The complementary problem of ``too many
answers'' has been previously addressed in the literature -- see, for
example \cite{carey-enough,carey-distance}.)

Two approaches, both implemented on the \emph{client-side}, have been
proposed for the ``too few answers" problem: The 64K Inc.\cite{64k}
engine augments query results (if any) with statistical information
about the underlying data distribution. Users are expected to utilize
this information to rephrase their queries appropriately. However, it
is unrealistic to expect that na\"{\i}ve Web users will be able (or
willing) to perform the calculations necessary to rephrase their
queries.

An alternative approach was proposed in Eureka~\cite{eureka}. In
response to the initial user query, Eureka caches the relevant portion of the
\emph{database} at the client machine, allowing follow-up exploratory
queries to be answered locally. A major drawback is that the user
needs to install a customized software for each of the Web services
that she wishes to access.  In addition, this strategy may not be feasible
for resource-constrained client devices which may be unable to host
the entire database segment, or which are connected through a
low-bandwidth network.

Finally, yet another possibility is to convert the user's range query
into a point query (\eg by replacing the box represented by the query
with its centerpoint) and then to use one of the several Top-K algorithms
available in the literature (\eg \cite{bruno02}) with respect to this
point. However, this approach is unacceptable since it runs the risk of
not providing all the results that are part of the original user query.
Further, as discussed later in this paper, closeness to a point may not
be equivalent to closeness to the query box.

\subsection*{The SAUNA Technique}
\begin{figure}
\centerline{\psfig{file=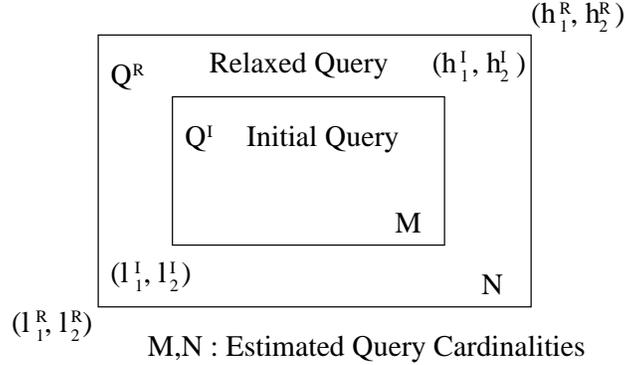,width=.5\textwidth}}
\caption{Range query relaxation in 2 dimensions}
\label{fig:relaxation}
\end{figure}

In this paper, we propose SAUNA (Stretch A User query to get N Answers),
a \emph{server-side} solution for efficiently supporting Web-based
exploratory queries.  More formally, given an initial user query $Q^I$
(which we expect to return $M$ answers), and given the desired number
of answers $N$, if $N > M$, SAUNA derives a new relaxed query $Q^R$
which \emph{contains} $Q^I$ and is expected to have $N$ answers.
A pictorial representation of a SAUNA relaxation is shown in
Figure~\ref{fig:relaxation} for a two-dimensional range query.

Note that a variety of relaxed queries, which may even be infinite
in number, could be derived that obey the above constraints. In
this solution space, SAUNA aims to deliver a relaxed query that (a)
minimizes the distance of the additional answers with respect to the
original query, that is, it aims to derive the closest $N-M$ answers,
and (b) minimizes the data processing required to produce this set of
answers. The first goal is predicated on defining a distance metric for
points lying outside the original query -- this issue is well understood
for \emph{point-queries}~\cite{bruno02} but not for the \emph{range} (or
\emph{box}) queries that we consider here. Therefore, SAUNA incorporates a
box-query-specific distance metric that is suitably weighted to produce
relaxed queries of a desired shape (\eg aspect-ratio preserving
with respect to the original query). To achieve the second goal,
SAUNA utilizes multi-dimensional histograms as the tool for query size
estimation. Histograms~\cite{dewitt-equidepth,poosala-avi,poosala-histograms} are the de facto standard technique for maintaining
statistical summaries in current database systems, and therefore our
system is easily portable to these platforms.  While uni-dimensional
histograms are currently the norm, techniques for easily building and
maintaining their multi-dimensional counterparts have recently appeared
in the literature~\cite{selftune}.

As we show in Section~\ref{sec:expt}, a detailed performance evaluation of
SAUNA over a variety of real and synthetic multi-dimensional data sets
stored on a Microsoft SQL Server 2000 engine indicates that its
relaxed queries can significantly reduce the costs associated with
exploratory query processing, and in fact, often compare favorably
with the optimal-sized relaxed query (obtained through off-line
processing). Further, these improvements are obtained even when
the memory budget for storing statistical information is extremely
limited.

\subsection*{Organization}
The remainder of this paper is organized as follows: The relaxation problem
is formally defined in Section~\ref{sec:problem}. Distance metrics for box queries are discussed in Section~\ref{sec:metrics}.  The SAUNA query relaxation strategy is
presented in Section~\ref{sec:strategies}.  The performance model and
the experimental results are highlighted in Section~\ref{sec:expt}.
Related work on query relaxation is reviewed in Section~\ref{sec:related}.
Finally, in Section~\ref{sec:concl}, we summarize the conclusions of
our study and outline future research avenues.

\section{Problem Definition}

\section{Distance Metrics for Box Queries}
\label{sec:metrics}

Most distance functions used in practice are based on the general
theory of \emph{vector p-norms}~\cite{Gradshteyn@book2000}, with $1
\leq p \leq \infty$.  For example, $p = 2$ gives the classical
Euclidean metric, $p = 1$ represents the Manhattan metric, and $p
=\infty$ results in the Max metric. In the remainder of this paper,
for ease of exposition, we assume that all distances are measured with
the Euclidean metric.  Note, however, that the SAUNA relaxation
algorithm can be easily adapted to any of the alternative metrics.

\subsection{Reference Points}
When computing the distances of database tuples with respect to
\emph{point} queries, it is clear that the distances are always to be
measured (whatever be the metric) between the pair of points represented
by the database tuple and the point query.  However, when we come to
\emph{box} (range) queries, which is the focus of this paper, the issue
is not so clear-cut since it is not obvious as to which point in the box
should be treated as the reference point. In fact, it is even possible
to think of distances being measured with respect to \emph{a set} of
reference points.

\begin{figure}
\centerline{\psfig{file=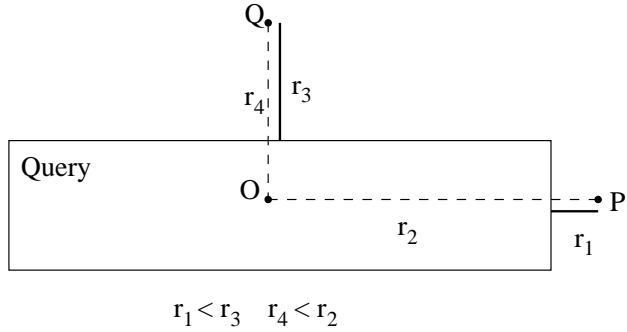,width=.5\textwidth}}
\caption{Measuring distance from periphery. $P$ is closer to periphery than $Q$}
\label{fig:compare-metrics-points}
\end{figure}

One obvious solution is to take some point inside the box (\eg the
center), treat the box as being represented by this point, and then
resort to the traditional distance measurement techniques.  However, this
formulation appears highly unsatisfactory since the spatial structure of
the box, which is representative of the user intentions, is completely
ignored. Instead, we contend here that the user's specification of a box
query implies that she would prefer answers that are \emph{close to the
periphery} of the box.  To motivate this, consider the example situation
shown in Figure \ref{fig:compare-metrics-points}, where point $P$ is
farther from the box center than point $Q$ \ie $r_2 >  r_4$, but $P$'s distance from the
closest face of the box is smaller than the corresponding distance for
$Q$ \ie $r_1 < r_3$. In this situation, we expect the user to prefer point $P$ over $Q$
since there is less deviation with respect to the complete box.

The above observation can be formally captured by the following reference
point assignment technique: For measuring the distance between a point
$P$ and a query box $B$, the reference point on $B$ is the point of
intersection of the perpendicular line drawn from $P$ to the nearest face or
corner of the box $B$.

We could, of course, have devised more complex reference point assignments
-- for example, compute the average of the distance between $P$ and all
corners of the box $B$, with the box corners operating as a universal set
of reference points.  However, we expect that the above simple formulation
may be sufficient to express the expectation of a significant fraction
of users of Web services, and further, more complex assignments can be
directly accommodated, if required, in the SAUNA relaxation algorithm.

In summary, given a point $P = \{p_1,p_2,...,p_D\}$ and a box-query $B$ with lower and upper limits $l_i(B)$ and $h_i(B)$ respectively, 
we denote the component of distance on 
the $i$-th dimension as
\begin{eqnarray} d_i(P,B) & = & p_i - h_i(B)  ~~~   if ~~ p_i > h_i(B) \nonumber \\
			     & = & l_i(B) - p_i ~~~~  if ~~~ p_i < l_i(B) \nonumber \\	
			     & = & 0 	~~~~~~~~~~~~~~~~~ otherwise	\nonumber
\label{eqn:dist-dmn}
\end{eqnarray}
and the overall (Euclidean) distance between $P$ and $B$ as
\begin{equation}
 dist(P,B) = \sqrt{\sum_{i=1}^{D} (d_i(P,B))^2} 
\label{eqn:distance}
\end{equation}
Note that with this formulation, all points that lie \emph{within} or
\emph{on} the box have an associated distance of zero.

\subsection{Attribute Weighting}
An implict assumption in the above discussion was that relaxation on all
dimensions was equivalent.  However, it is quite likely that the user
finds relaxation on some attributes more desirable than on others. For
example, a business traveler may be time-conscious as compared to price,
whereas a vacationer may have the opposite disposition.  Therefore,
we need to \emph{weight} the distance on each dimension appropriately.
That is, we modify Equation~\ref{eqn:distance} to
\begin{equation}
 dist(P,B) = \sqrt{\sum_{i=1}^{D} (d_i(P,B) * w_i)^2} 
\label{eqn:weight}
\end{equation}
\noindent
where $w_i, \;w_i \geq 0$ is the weight assigned to dimension $i$.

One option certainly is to explicitly acquire these weights from the user,
and use them in the above equation.
However, as a default in the absence of these inputs, we can resort to the
following: \emph{Use the box shape as an indicator of the user's intentions}.
Specifically, we can assume that the user is willing to accept a
relaxation on each range dimension that is \emph{proportional} to the
range size in that dimension, \ie  the user would
prefer what we term as an \emph{Aspect-Ratio-Preserving} relaxation.
This metric preserves the aspect ratio of user-supplied query hence the name.
This objective can be easily implemented by setting
\begin{equation}
w_i^{aspect} = \frac{1}{Asp\_ratio(i)} =  \frac{Max^D_{i=1} (h_i(B) - l_i(B))}{h_i(B) - l_i(B)}
\label{eqn:asp}
\end{equation}

An alternative interpretation of the user's box-query structure could be
that attributes should be relaxed in \emph{inverse} proportion to their
range sizes, since the user has already \emph{built-in} relaxation into
the larger ranges of her query.  This can be implemented with the following
distance function
\begin{displaymath}
w_i^{inverse} = 
Asp\_ratio(i) = \frac{h_i(B) - l_i(B)}{Max^D_{i=1} (h_i(B) - l_i(B))}
\label{eqn:inverse}
\end{displaymath}

\begin{figure}
\centerline{\epsfig{file=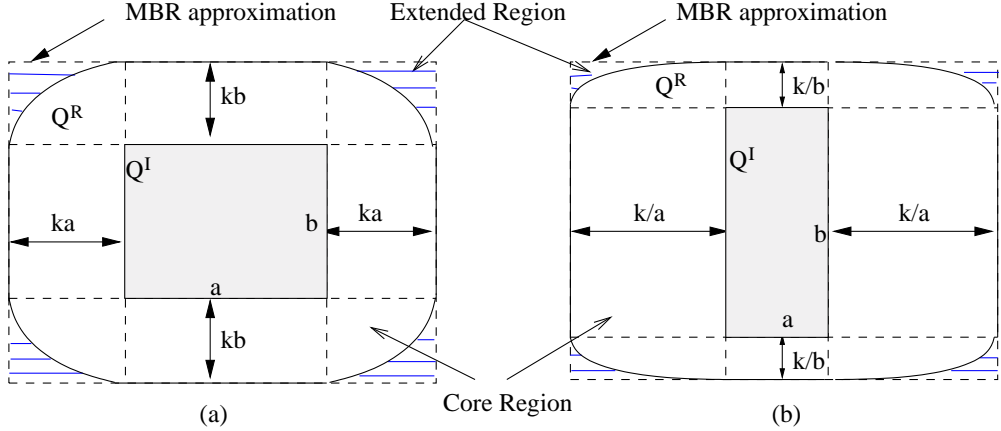,width=.8\textwidth}}
\caption{Distance Metrics and Relaxation regions: (a) Aspect (b) Inverse }
\label{fig:distance-metrics}
\end{figure}

Figure~\ref{fig:distance-metrics} shows an example of the relaxed
queries produced by using the \emph{Aspect} and \emph{Inverse}
metrics, respectively. Given a constant $k$ and relaxation units 
$a$ and $b$ (in the $x$ and $y$ axes, respectively), we see in these figures that the locus of points
equidistant from the original query is not hyper-rectangular in
the corners.  Since relational databases can execute only hyper-rectangular
queries, we approximate the relaxed queries by their \emph{Minimum
Bounding (Hyper)-Rectangles} (\emph{MBR}s).  We refer to the area enclosed
within the locus as the \emph{core} region and the area between the
\emph{core} region and the MBR rectangle as the \emph{extended region}.

If our goal is to produce the \emph{closest set} of answers to the
query box, then we need to explicitly prune the extended region points.
This is because there may be a point lying just outside the current
MBR whose distance is less than that of the extended region points.
We term this as a \emph{distance-preserving} relaxation. However, if minor
deviations from the optimal set of answers is acceptable, then we can
settle for a \emph{box-preserving} relaxation instead, wherein answers from
the extended region are also included in the answer set.  Our experimental
results indicate little performance difference for these alternative
relaxations -- therefore, we assume a box-preserving relaxation in the
remainder of this paper.

As a final point, note that if the user has specified a point query
as opposed to a box query, then the above formulation degenerates to a
traditional Top-N query~\cite{bruno02}, where the goal is to find the
nearest N neighbors to the query point.
\section{The SAUNA Relaxation Algorithm}	
\label{sec:strategies}

We propose SAUNA, a simple query relaxation mechanism that attempts to
ensure the desired cardinality and quality of answers while simultaneously
trying to reduce the cost of relaxed query execution.  Specifically,
our algorithm generalizes to box queries the approach taken for point
queries in \cite{bruno02,surajit99}.

Our relaxation strategy leverages histograms for query size estimation.
Histograms are the de facto standard technique for maintaining
statistical summaries in current database systems, and therefore
SAUNA is easily portable to these platforms.  In particular, we use
multi-dimensional histograms for the experiments reported in this
study. Although multi-dimensional histograms have been touted as being
resource-intensive to create and maintain, recent work~\cite{selftune}
has addressed this problem by proposing an online adaptive mechanism
for easily building and maintaining multi-dimensional histograms, the
so-called self-tuning histograms.

Due to their summary nature, histograms can provide only estimates, and
not the exact values. Therefore, when relaxing a query to produce $N$
answers, there is always a risk of either under-estimation or
over-estimation of the cardinality of the answer set. While
under-estimation results in inefficiency due to accessing more database
tuples than necessary, over-estimation requires the query to be relaxed
further and submitted again -- a \emph{restart} in the terminology of 
\cite{surajit99}.

Estimation strategies possible in this environment include a
conservative approach that completely eliminates restarts at the risk of getting
many more tuples than necessary, and an optimistic approach that trades
restarts for improved efficiency. These \emph{No-Restarts} and \emph{Restarts}
approaches were implemented in \cite{surajit99} by assuming that all
database tuples in a histogram bucket are at the maximum or minimum
distance, respectively, with respect to the point query.  Note that 
for a point query, there is always a unique location on a
histogram bucket which is at a minimum (maximum) distance from the
point query. However, when we consider box-queries in conjunction with the
periphery-based distance metric described in the previous section,
there is a \emph{set} of points on the histogram bucket that are all
at the same minimum (maximum) distance from the box query.
In Figure~\ref{fig:Dist}, we present
the MinDist and MaxDist algorithms to find these minimum and maximum distances,
respectively. Both
these algorithms are linear in the number of query attribute dimensions.
We describe below the various relaxation strategies for box queries that
are based on these distance computations.

\subsection{Box-Restarts Strategy}
In this approach, all tuples inside a histogram bucket are assumed to
be present on a locus of \emph{minimum} distance from the query box.
Since both the query box and the histogram bucket are $D$-dimensional
hyper-rectangles, the minimum distance between them is the minimum
distance between any pair of their $D-1$ dimensional hyper-rectangle
surfaces. We use the \emph{MinDist} algorithm (Figure~\ref{fig:Dist}(a))
to compute this minimum distance.  MinDist locates one of the points at
minimum distance on the bucket and then computes the distance of that
point from the query box.  In the algorithm, $b^l_i$ and $b^h_i$ are
the lower and upper bounds of the bucket in the \emph{i}-th dimension,
while $q^l_i$ and $q^h_i$ are the corresponding lower and upper bounds
of the box query.  It should be noted here that the identification of
the nearest point in the \emph{MinDist} algorithm is \emph{independent}
of the specific distance metric (including attribute weighting) chosen
for computing the minimum distance.



\begin{figure}
\begin{center}
\begin{footnotesize}
\vsp{0.9} 
\begin{tabular}{|c|c|} \hline
\begin{minipage}{2.8in}
\begin{tabbing} \ \= \ \ \= \ \ \ \ \ \ \ \ \= \kill
$Algorithm ~ \emph{MinDist}~ (Box~q, Bucket~b,$ \\
\> \> \> $~Metric~ metric)$ \{ \\
\> $Point ~  Nearest,~Nearest^l,~Nearest^h$; \\	
\> $\forall~i~:~ 1 ~ \leq ~ i ~\leq ~ D ~$ \\
\> $ begin$  \\
\> $Nearest^l_i ~ = ~ q^l_i ~~~~~~ if ~ b^l_i ~ \leq q^l_i ~ \leq ~b^h_i $ \\
\> \> $~ = ~ b^l_i ~~~~~~ if ~ q^l_i ~ < ~ b^l_i$ \\
\> \> $~ = ~ b^h_i ~~~~~~ otherwise$ \\ 
\> $Nearest^h_i ~ = ~ q^h_i ~~~~~~ if ~ b^l_i ~ \leq q^h_i ~ \leq ~b^h_i $ \\
\> \> $~ = ~ b^l_i ~~~~~~ if ~ q^h_i ~ < ~ b^l_i$ \\
\> \> $~ = ~ b^h_i ~~~~~~ otherwise$ \\ 
\> $ ~if ~| q^l_i ~ - ~ Nearest^l_i| ~ < ~ | q^h_i ~ - ~ Nearest^h_i|$ \\
\> \> $Nearest_i  ~  = ~ Nearest^l_i$ \\
\> $else $ \\
\> \> $Nearest_i     ~  = ~ Nearest^h_i ~~~~~~ $ \\
\> $ end ~ \forall ~ i$ \\
\> $return ~ dist_{metric}~(Nearest,~q)$	\\
 		\} 
\end{tabbing}
\end{minipage}
& 
\begin{minipage}{2.8in}
\begin{tabbing} \ \= \ \ \= \ \ \ \ \ \ \ \ \= \kill
$Algorithm ~ \emph{MaxDist}~ (Box~q, Bucket~b,$ \\
\> \> \> $~Metric~ metric)$ \{ \\
\> $Point ~  Farthest,~Farthest^l,~Farthest^h$; \\	
\> $\forall~i~:~ 1 ~ \leq ~ i ~\leq ~ D ~$ \\
\> $ begin$ \\
\> $Farthest^l_i ~ = ~ b^l_i ~~~~~~~ if ~ q^l_i ~ \leq b^l_i $ \\
\> \> $~~ = ~ b^h_i ~~~~~~~ otherwise$ \\ 
\> $Farthest^h_i ~ = ~ b^l_i ~~~~~~~ if ~ q^h_i ~ \leq b^l_i$\\
\> \> $~~~ = ~ b^h_i ~~~~~~~ otherwise$ \\ 
\> $ ~if ~| q^l_i ~ - ~ Farthest^l_i| ~ > ~ | q^h_i ~ - ~ Farthest^h_i|$ \\
\> \> $Farthest_i  ~  = ~ Farthest^l_i$ \\
\> $~ else $ \\
\> \> $~ Farthest_i  = ~ Farthest^h_i ~~~~~ $ \\
\> $ end ~ \forall ~ i$ \\
\> $return ~ dist_{metric}~(Farthest,~q)$	\\
 		\} \\
\end{tabbing}
\end{minipage} \\ 
(a) MinDist  & (b) MaxDist \\ \hline

\end{tabular}
\caption{Algorithms for computing distances}
\label{fig:Dist}
\defspacing
\end{footnotesize}
\end{center}
\end{figure}

In the Box-Restarts relaxation strategy, we compute the minimum distances
of all histogram buckets from the query box, and then sort these buckets
in increasing order of these distances. We assume that relaxing the query
up to the minimum distance of some bucket implies that the relaxed query
includes all tuples in that bucket. Hence we choose the largest distance
from the set of bucket distances such that the relaxed query is expected
to contain $N$ tuples. Since the underlying assumption that all points
in a bucket are as close as possible to the query box is optimistic,
the Box-Restarts strategy does not guarantee that the relaxed query
will indeed return $N$ tuples.

\subsection{Box-NoRestarts Strategy}
In this approach, all tuples inside a histogram bucket are assumed to
be present on a locus of \emph{maximum} distance from the query box. We
use the \emph{MaxDist} algorithm (Figure~\ref{fig:Dist}(b))
to compute this maximum distance.  The process we follow for finding
the Box-NoRestarts relaxation distance is the same as that for the
Box-Restarts approach outlined above. Since the relaxed query is
guaranteed to cover all the histogram buckets at a distance less than
or equal to relaxation distance, the Box-NoRestarts strategy guarantees
that the relaxed query shall return at least $N$ answers. This guarantee
is obtained at the cost of efficiency in that many more tuples than
strictly necessary may have to be processed to find the desired
answer set.

\begin{figure}
\centerline{\epsfig{file=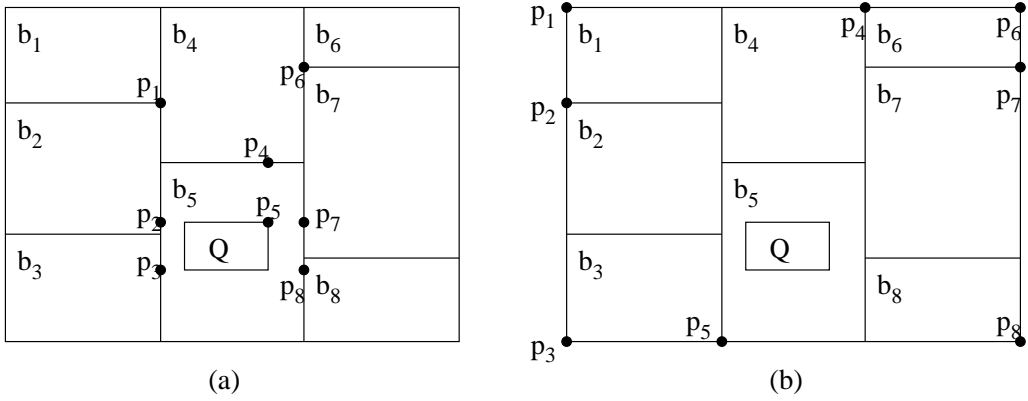}}
\caption{Box query relaxation strategies. 
(a) \emph{Box-Restarts} (b) \emph{Box-NoRestarts}.} 
\label{fig:box-method}
\end{figure}

To make the above discussion concrete, sample points chosen by the MinDist
and MaxDist algorithms are shown in Figures~\ref{fig:box-method}(a)
and ~\ref{fig:box-method}(b), respectively.  In these figures, $Q$
is the query box, $b_1$ through $b_8$ are the histogram buckets in the
2-dimensional space, and $p_1$ through $p_8$ are the points chosen by
the algorithms.
Note that while minimum distance points can be located on the query
box itself (\eg $p_5$ in Figure~\ref{fig:box-method}(a)), the maximum
distance points always have to be on the corners of the histogram
bucket (all $p_i$ in Figure~\ref{fig:box-method}(b)).

\subsection{Box-Dynamic Strategy} \label{sec:bdyn}
Since Box-Restarts and Box-NoRestarts represent extreme solutions, an
obvious question is whether an intermediate solution that provides the
best of both worlds can be devised?  For this, we adopt the dynamic
workload-based mapping strategy of \cite{bruno02}, which attempts
to find the relaxation distance that minimizes the expected number of
tuples retrieved for a set of queries while ensuring a reduced number of
restarts. This is implemented as follows: Given $\alpha$ as a parameter
such that
\begin{displaymath}
	d_q(\alpha)= d^{BR}_q ~ + ~ \alpha  \left(d^{BNR}_q - d^{BR}_q \right)
\end{displaymath}
where $d^{BR}_q$ and $d^{BNR}_q$ are the \emph{Box-Restarts} and
\emph{Box-NoRestarts} distances for query $q$, we need to dynamically
find the value of $d_q(\alpha)$ that minimizes the average number of
tuples retrieved for a given query workload.  Since $d_q(\alpha)$ is a
unidimensional function of $\alpha$, the \emph{golden search} algorithm
\cite{golden-search} can be utilized to estimate this optimal value of
$\alpha$. Note that this approach requires an initial ``training workload"
to determine a suitable value of $\alpha$, which can then be used in the
subsequent ``production workloads".

In the remainder of this paper, we present results only for the
Box-Dynamic strategy since we found that it consistently
outperformed the extreme strategies.

\subsection{Relaxation Algorithm}
\label{subsec:relax-algo}
While the Box-Dynamic strategy does reduce the likelihood of restarts,
it does not completely eliminate them.  To ensure that we do not get
into a situation where there are repeated restarts of a given query,
we follow the strategy that if the Box-Dynamic strategy happens to fail
for a particular query, then we immediately resort to the conservative
Box-NoRestarts strategy -- that is, all queries are relaxed with at
most one restart.  The complete set of steps of the SAUNA relaxation
algorithm is shown in Figure~\ref{fig:relaxation-alg}.

\begin{figure}
\begin{center}
\begin{tabular}{|l|}
\hline
$Algorithm ~ \emph{SAUNA Relaxation}~ (Query~Q^I,Integer N)$ \\
		\{\\
1\hspace*{1mm}		$M$ = estimateCardinality($Q^I$); \\
2\hspace*{1mm}		if $M < N$ \\
3\hspace*{7mm} 	           $Q^R$ = relaxBoxDynamic($Q^I$); \\
4\hspace*{7mm}             numAnswers = execute($Q^R$);\\
5\hspace*{7mm} 		   if numAnswers $\geq N$ return the $N$ nearest answers; \\
6\hspace*{7mm}		   else \\
7\hspace*{10mm}		      $Q^{R'}$ = relaxNoRestart($Q^I$); \\
8\hspace*{10mm}	      execute($Q^{R'}$); \\
9\hspace*{1mm}           else \\
10\hspace*{7mm}		      numAnswers = execute($Q^I$); \\
11\hspace*{7mm}                if numAnswers $\geq N$ return all answers; \\
12\hspace*{7mm}		      else \\
13\hspace*{10mm}		         $M$ = numAnswers; \\
14\hspace*{10mm}	                 go to Step 3; \\
 		\} \\
\hline
\end{tabular} \\
\caption{SAUNA relaxation algorithm}
\label{fig:relaxation-alg}
\end{center}
\end{figure}


\subsection{Handling Categorical Attributes}
\label{subsec:categorical}
An implicit assumption in the discussion so far was that all attributes
are either continuous or discrete with inherent ordering among the
values. In practice, however, some of the dimensions may be \emph{categorical} in
nature (\eg color in an automobile database), without a natural ordering
scheme. We now discuss how to integrate categorical attributes into
our relaxation algorithm. 

In the prior literature, we are aware of two techniques that address 
the problem of clustering in categorical spaces -- the first approach 
is based on \emph{similarity}~\cite{categorical} and the second is based on 
\emph{summaries}\cite{CAC}. While both techniques can be used in our 
framework to calculate distances, we restrict our attention to the 
former in this paper.

The similarity approach works as follows:
Greater weight is given to ``uncommon feature-value matches'' in
similarity computations. For example, consider a categorical attribute
whose domain has two possible values, $a$ and $b$.  Let $a$ occur more
frequently than $b$ in the dataset. Further, let $i$ and $j$ be tuples
in the database that contain $a$, and let $p$ and $q$ be tuples that
contain $b$.  Then the pair $p,q$ is considered to be more similar than
the pair $i,j$, \ie $Sim(p, q) > Sim(i, j)$;  in essence, tuples that 
match on less frequent values are considered more similar.  

Quantitatively, similarity values are normalized to the range [0,1].
The similarity is zero if two tuples have different values for the
categorical attribute.  If they have the same value $v$, then the similarity
is computed as follows: 
\[
Sim(v) = 1 - \sum _{l \in MoreSim(v)} \frac {f_l(f_l-1)}{n(n-1)}\\
\]
where $f_l$ is frequency of occurrence of value $l$, $n$ is the number
of tuples in the database, and MoreSim(v) is the set of all values in the
categorical attribute domain that are more similar or equally similar as
the value $v$ (\ie they have smaller frequencies).

We cannot directly use the above in our framework since our
goal is to measure \emph{distance}, not similarity. At first glance,
the obvious choice might seem to be to set \emph{distance} $=1-$
\emph{similarity}.  But this has two problems: Firstly, tuples with
different values in the categorical attribute will have a distance
of 1.  Secondly, tuples with identical values will have a non-zero
distance. Both these contradict our basic intuition of distance.

Therefore, we set the definition of distance as follows: If two tuples
have the same attribute value, then their distance is zero. Tuples
with different values will have distances based on the frequencies of
their attribute values. The more frequent the values, the less is the
distance. For example, if the categorical attribute has values $a$,
$b$ and $c$ in decreasing order of frequencies, $DIST(a,c) <
DIST(b,c)$, since $a$ is more frequent than $b$. In general, given
tuples with values $v_1$ and $v_2$, we can quantitatively define 
\begin{eqnarray*}
DIST({v_1,v_2}) & = & Sim(v_1) * Sim(v_2) \; \; \mbox{if} \; \; v_1 \neq v_2 \\
& = & 0 \; \; \; \; \; \; \;  otherwise
\end{eqnarray*}

\section{Experimental Results}
\label{sec:expt}

\renewcommand{\topfraction}{.9}
\renewcommand{\bottomfraction}{.9}
\renewcommand{\floatpagefraction}{1}
\renewcommand{\textfraction}{0.1}

\subsection{Experimental Settings}
We used a variety of synthetic and real-world data sets to evaluate
SAUNA -- these datasets are the same as those used in \cite{bruno02}. 
The real-world data sets consisted of the US census data set
($199,523$ tuples) and the Forest data set ($581,012$ tuples) obtained
from~\cite{www:datasets}. We selected from these data sets the same
set of attributes as~\cite{bruno02}.
The synthetic data consisted of the \emph{Gauss} and \emph{Array} data
sets, each containing $500,000$ tuples. The \emph{Gauss} data
sets~\cite{golden-search} were generated using predetermined number of
overlapping multidimensional gaussian bells. Each bell was
parameterized by the variance and zipfian parameter.  The \emph{Array}
data sets were generated using zipfian distribution~\cite{zipf-book} 
for frequency of data values along each attribute.  The value sets of
each attribute were generated independently. The values of zipfian
parameter for both these data sets were chosen to be $0.5, 1, 1.5$ and
$2$.

All the experiments were performed using multidimensional equi-depth
histograms~\cite{dewitt-equidepth}, as they are both accurate and
simple to implement.
Further, an $N$-dimensional unclustered  
concatenated-key $B^+$-tree multidimensional index covering all
the query attributes was built over each data set.

The query workload consists of queries with the number of range
dimensions varying from 2 to 4, which is typical of many Web applications.
The specific queries were generated by moving a query template over the
entire domain space. The size of each query template was assigned so as
to ensure that the complete domain space was covered by a hundred queries.
This query density was sufficient to ensure that most queries suffered
from the problem of too few answers and therefore required relaxation.
All results we report are averages for this set of hundred queries.

Besides different datasets, we also evaluated the performance of SAUNA
with respect to  (a) varying the number of buckets in the histogram;
(b) varying $N$, the desired result cardinality;
(c) varying the skew in the data; and,
(d) varying the distance metric.
To serve as comparative yardsticks for SAUNA's performance, we used two
benchmarks: 

\begin{description}
\item[Sequential (SEQ)]: In this strategy, a sequential scan of the database
is made in order to produce a sorted list of the tuples w.r.t. their 
distance from the query box, after which the top $N$ tuples are returned. 
\item[Optimal (OPT)]: This strategy refers to a hypothetical optimal relaxation
strategy which produces the \emph{minimally relaxed query} that contains
the desired answer set.   Note that the minimum bounding hyper-rectangle
enclosing the $N$ nearest tuples of a query box is not guaranteed
to return $N$ answers only and often returns more than $N$ answers.
Further, it is not possible for
any relaxation technique, without observing the actual data tuples, to
retrieve tuples equal to OPT tuples. In our
experiments, the answers for OPT were found through an offline complete
scan of all the data tuples.
\end{description}

In the following experimental descriptions, the labels in the X-axis
of the graphs are of the form $Dataset(Dim)Strat$, where: $Dataset$
is the name of the dataset -- ``cen'' refers to the Census data sets,
``gz'' refers to the \emph{Gauss} data sets, ``arr'' refers to the
\emph{Array} data sets, and ``cov'' refers to the Forest data set; $Dim$
is the number of dimensions of the data; and $Strat$ is the relaxation
strategy -- where ``opt'' indicates the optimal values and ``B-dyn''
refers to the \emph{Box-dynamic} strategy.  For all the results, unless
otherwise mentioned, the default settings were zipfian parameter $z =
1$, number of dimensions $= 3$, number of desired answers $N = 10$,
\emph{Aspect} distance metric and number of histogram buckets $= 256$.
Finally, the \emph{Box-dynamic} strategy (see Section~\ref{sec:bdyn})
is used for SAUNA relaxation in all the experiments presented here.
Our experiments were conducted on a Pentium IV  machine 
running the Windows 2000 operating system.


\subsection{Experiment 1: Basic SAUNA performance} 

\begin{figure*}
\centerline{\psfig{file=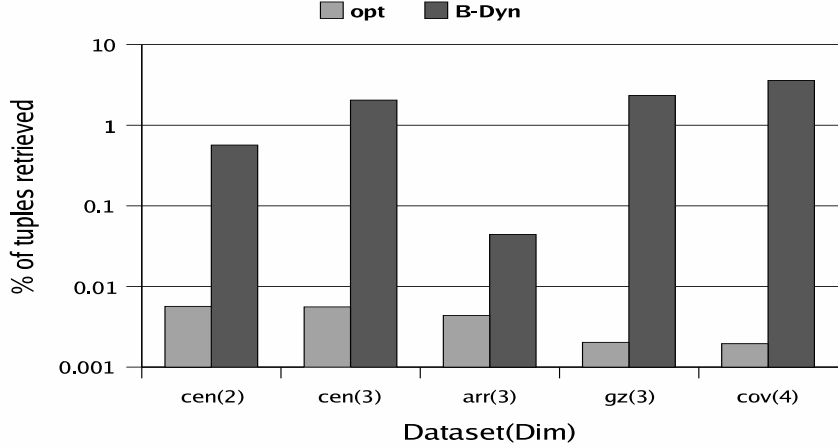,height=2.5in,width=.7\textwidth}}
\caption{Percentage  of tuples retrieved}
\label{graph:datasets}
\end{figure*}

The performance of SAUNA and OPT on the various datasets for the default
parameter settings is shown in Figure~\ref{graph:datasets} with respect
to the number of tuples retrieved (note that the Y-axis is shown on a \emph{log scale}).
The first point to observe here is that for all the datasets, SAUNA requires
processing less than 4\% of the tuples -- in fact, for the \emph{census}
and \emph{array} datasets they are less than $1\%$.  Secondly, note
that there is quite a substantial gap between the optimal performance
and that of SAUNA. This is due to the fact that SAUNA has to depend on
statistical information that is limited by a tight memory budget (only
256 histogram buckets, consuming around 5KB memory, were used in this experiment).

%
%

\begin{figure}
\centerline{\psfig{file=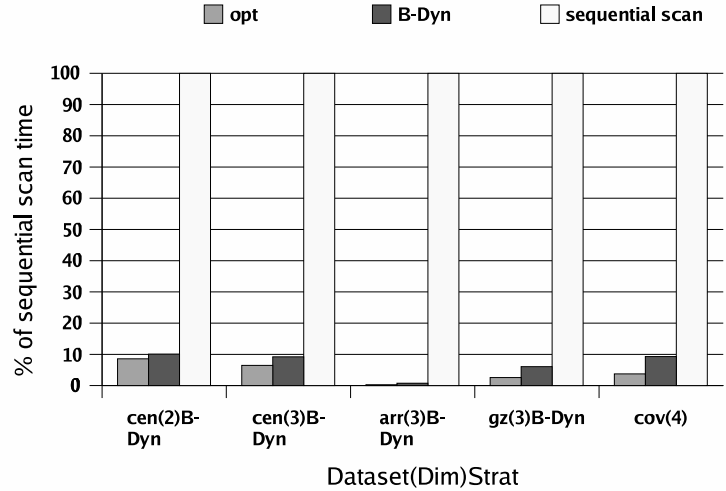,width=.6\textwidth}}
\caption{Execution time of SAUNA relative to SEQ}
\label{graph:time}
\end{figure}

In  Figure~\ref{graph:time}, we show the running times of SAUNA and
OPT strategy(excluding the time required to find the optimal
relaxed query), normalized to the execution time of SEQ, for the various datasets.
The first point to note here is that the SAUNA execution times are below
$10\%$ of the sequential scan time for all the datasets. Secondly, for
the \emph{census} and \emph{array} datasets the SAUNA times are close
to that of OPT, and even for the other datasets the difference is small.

The execution time figures clearly indicate the efficiency of SAUNA
w.r.t. the optimal strategy. Again, it should be noted that it is not
the relaxation algorithm, but the quality of the histograms (the type
and number of buckets) that affect the efficiency of SAUNA as compared
to the optimal in terms of number of tuples retrieved or the execution
time. By increasing number of histogram buckets, we expect that SAUNA
would perform closer to the optimal.

Finally, in order to illustrate the usefulness of SAUNA, we compared it
against a model of a \emph{user's} (manual) relaxation session. Although
each  user is likely to have her own relaxation strategy, we choose
the following simple model to serve as an indicator of the potential of
automated relaxation: The user's performance is measured as the total
execution time of a sequence of queries, beginning with the original
query, which is relaxed by $20\%$ in each dimension ($10\%$ in each
direction) after each iteration 
until the optimal relaxation is achieved.

When the execution time of SAUNA was compared against the above MANUAL
procedure, we obtained Table~\ref{table:manual}.  We see in this
table that SAUNA outperforms the MANUAL approach by a substantial
factor time-wise. Note further that these numbers are extremely
\emph{conservative} since they completely ignore the high 
latencies (\ie from the network, Web server and back-end database)
that are involved in submitting a sequence of queries.  Finally,
observe that even though the ranges of the queries were as large as
$0.1\%$ of the domain in each dimension,  they required numerous
iterations to reach the desired value. It is likely that users faced
with such situations would prematurely terminate their interaction
with the Web service out of frustration.

\begin{table}
\begin{center}
\begin{tabular}[h]{|c|c|c|c|c|c|c}
\hline
Dataset	             & cen(2) & cen(3) & arr(3) & gz(3) & cov(4)  \\ \hline
Manual Iterations    & 112    &  129   &  23    & 116   &  16     \\
ManualTime/SaunaTime & 11     &    9   &  1.2   & 3.5   &  2      \\
\hline
\end{tabular}
\end{center}
\caption{SAUNA versus MANUAL relaxation}
\label{table:manual}
\end{table}

\subsection{Experiment 2: Varying Number of Histogram Buckets}

In this experiment, we investigated the performance improvements
that could be obtained if our tight memory budget for statistical
information was somewhat relaxed. In particular, we varied the
memory budget from the default 5 KB to about 100 KB.

\begin{figure}
\centerline{\psfig{file=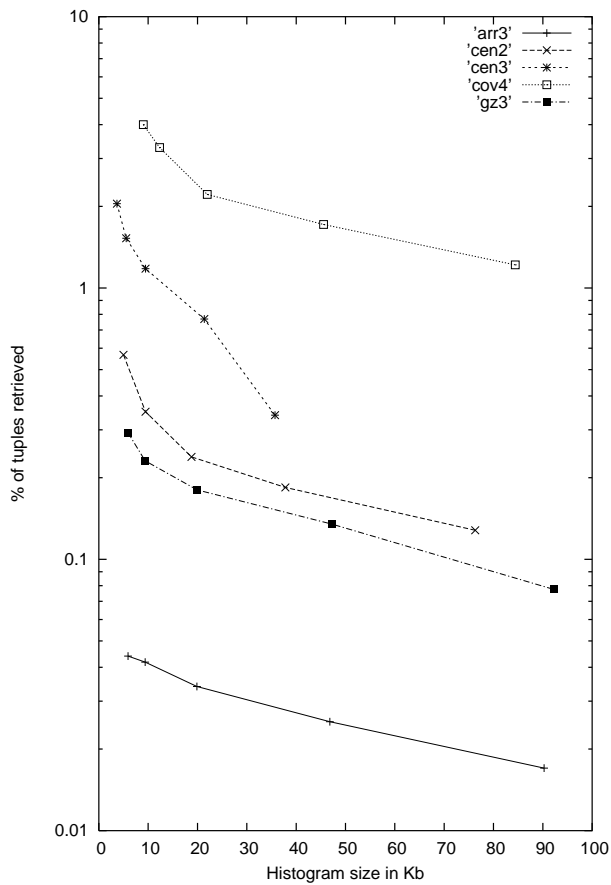, width=.5\textwidth}}
\caption{ Varying Number of Histogram Buckets}
\label{graph:buckets}
\end{figure}

The results of this experiment are shown in Figure~\ref{graph:buckets}.
It can be clearly seen here that the number of tuples retrieved approaches
optimal values with increasing number of buckets. This supports our
claim that SAUNA is limited by the quality of histogram statistics only.

\subsection{Experiment 3: Varying N}
%

\begin{figure*}
\centerline{\psfig{file=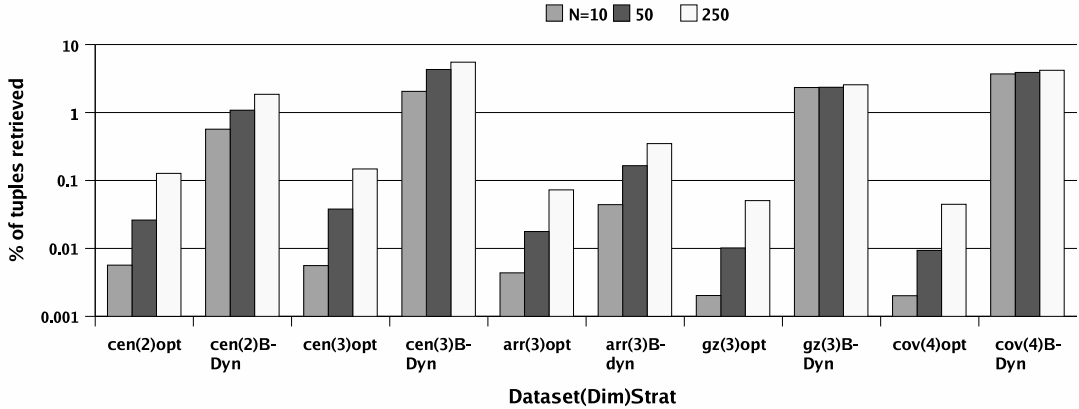,width=.9\textwidth}}
\caption{Percentage  of tuples retrieved: Varying $N$}
\label{graph:k}
\end{figure*}

We now move on to evaluating the effect of the choice of $N$,
the desired answer cardinality, on the performance of SAUNA.
The performance for values of $N = 10, 50, 250$  is shown in
Figure~\ref{graph:k}.  We see here that, in most cases, the cost does not increase
considerably with increasing values of $N$. This is because as
$N$ increases, the effective accuracy of the histogram becomes better
and better, and therefore there is lesser wasted effort.

The problems of dealing with \emph{too many} and \emph{too few} answers
have been addressed in many different contexts.  In the information
retrieval literature, various techniques have been proposed to both
relax and constrain keyword-based queries (see \eg
\cite{baeza-yates@book1999}). Many proposals for dealing with these problems
for more structured queries can be found in the database
literature~\cite{gaasterland@jiis1992,motro@acmis1988,carey-enough,
carey-distance,fagin,surajit99,bruno02}.

Recently, significant attention has been devoted to the evaluation of
\topN queries. \topN queries arise in many applications where users
are willing to accept non-exact matches that are close to
their specification. The answers to such queries consists of a ranked set of the
\emph{N} tuples in the database that best match the selection condition.
When a query returns too many answers, an interesting problem is how
to avoid processing data that will not contribute to the final \topN
results. Carey and Kossmann~\cite{carey-enough,carey-distance} proposed new
operators to improve the efficiency of \emph{Stop-After N} type
queries.
Fagin~\cite{fagin} addressed the problem of finding
\topN matches for queries that combine information from multiple
systems that may have different semantics.


Chaudhuri et al~\cite{surajit99} discuss the problem of evaluating
\topN equality selection queries that return too few answers.
They propose distance metrics for equality selection queries and
present histogram-based query relaxation strategies to automatically
relax such queries and return the desired number of answers.  They carry
forward their work in~\cite{bruno02}, where they introduce a dynamic
workload-aware strategy for processing \topN equality queries.  Their
work differs from ours essentially in the type of queries they support
-- whereas their work is limited to equality selection queries, SAUNA
supports the more general class of range queries.  Chen and
Ling~\cite{chen-ling} handle the same problem as \cite{surajit99}, but
using sampling as an estimation technique. They show that, unlike
histograms, sampling is quite efficient and effective when the
number of dimensions is large.

\section{Conclusions}
\label{sec:concl}

In this paper, we proposed SAUNA, a novel server-based framework for
automated query relaxation that improves the efficiency and efficacy
of query exploration over large and unknown data spaces. Unlike
previous approaches that are limited to point queries, SAUNA is able
to relax multi-dimensional range queries. Through the use of an
intuitive range-query-specific distance metric, SAUNA returns
high-quality answers that are \emph{closest} to the user-specified
query box. In addition, since histograms are used for query size
estimation, the SAUNA framework can be easily integrated with
commercial RDBMS that support histograms.  We also showed how
categorical attributes can be naturally integrated into this
framework.

Our experimental results indicate that SAUNA significantly reduces the
costs associated with exploratory query processing, and in fact, often
compare favorably with the optimal-sized relaxed query (obtained through
off-line processing). Further, these improvements are obtained even
when the memory budget for storing statistical information is extremely
limited.  Specifically, we found that even with as low a memory budget
as 5 KB, SAUNA was able to provide satisfactory relaxation retrieving
less than 10\% of the tuples in the database and taking less than 10\%
of the time taken by SEQ.  We also showed how it provides significant
benefits of up to an order of magnitude in execution time as compared
to user-driven manual relaxation.

There are two main directions we intend to pursue in future work:

\begin{itemize}
\item Since SAUNA relies on query cardinality estimations to perform
relaxation, its effectiveness is highly dependent on the estimation
mechanism.  Although the current implementation uses multi-dimensional
equi-depth histograms, we would like to experiment with other
strategies, \eg
\cite{poosala-histograms,poosala-avi,matias-wavelets}. 

\item Currently, when a restart is required, relaxation is applied and the
new relaxed query is executed. Note that this leads to redundant work,
as all answers for previous query are derived again. For future
work, we intend to investigate query splitting techniques (see
\eg~\cite{tan@icde1999}) to try and execute only the \emph{difference}
query.
\end{itemize}

\bibliographystyle{abbrv}

{\small
\baselineskip 10pt

\bibliography{dasfaa}
}

\end{document}